# Atomic perspective on the topological magnetism in kagome metal Co$_3$Sn$_2$S$_2$

Guowei Liu[1,5,*], Wei Song[1,*], Titus Neupert[2], M. Zahid Hasan[3,4], Hanbin Deng[1,†], Jia-Xin Yin[1,5,†]

[1]Department of Physics, State key laboratory of quantum functional materials, and Guangdong Basic Research Center of Excellence for Quantum Science, Southern University of Science and Technology, Shenzhen 518055, China.
[2]Department of Physics, University of Zurich, Zurich, Switzerland.
[3]Laboratory for Topological Quantum Matter and Advanced Spectroscopy, Department of Physics, Princeton University, Princeton, NJ, USA.
[4]Quantum Science Center at ORNL, Oak Ridge, TN, USA.
[5]Quantum Science Center of Guangdong-Hong Kong-Macao Greater Bay Area (Guangdong), Shenzhen, China.

*These authors contributed equally to this work.
†Corresponding authors. E-mail: yinjx@sustech.edu.cn, denghb@sustech.edu.cn

**Abstract**
Topological quantum materials with kagome lattices have attracted intense interest due to their unconventional electronic structures, which exhibit nontrivial topology, anomalous magnetism, and electronic correlations. Among these, Co$_3$Sn$_2$S$_2$ stands out as a prototypical kagome metal, uniquely combining intrinsic ferromagnetism with topologically nontrivial electronic states. This perspective presents a systematic overview of recent advances in studying kagome metal Co$_3$Sn$_2$S$_2$ achieved through scanning tunneling microscopy. We begin by introducing different methodologies for surface identification and propose using designer layer-selective chemical markers for conclusive surface identification. We then discuss the Berry curvature induced flat band orbital magnetism and the associated unconventional Zeeman effect. Furthermore, we explore boundary states arising from Weyl topology and analyze challenges in detecting Fermi arcs via quasiparticle interference patterns and in uncovering the topological aspect of the edge states. Finally, we review recent observations of spin-orbit-coupled quantum impurity states through spin-polarized tunneling spectroscopy, as well as their connection to Weyl topology and flat band magnetism. We also provide in-depth analysis and constructive comments on the limitations of the current research approach. This review highlights the critical role of scanning tunneling microscopy in unraveling the intricate interplay between topology, magnetism, and correlations at the atomic scale, and the methodology discussed here can be applied to study other topological quantum materials in general.

## 1 Introduction

Topological quantum materials have garnered extensive attention in condensed matter physics owing to their unconventional electronic states and promising prospects for spintronic and quantum computing applications[1-11]. Among them, kagome metals have emerged as ideal platforms for exploring a range of emergent phenomena, embracing both topology and correlation[12-34]. The kagome lattice is made of corner-sharing triangles. Owing to this unique lattice geometry, its electronic structure naturally hosts Dirac fermions, flat bands, and van Hove singularities[35-46]. The Dirac fermions encode band topology, flat band enhances electron correlations, and van Hove singularities can lead to Fermi surface instabilities. Their further interaction with spin magnetism drives the distinct emergent phenomena. Within this category, Co$_3$Sn$_2$S$_2$ has established itself as a prototypical kagome metal, uniquely combining intrinsic ferromagnetism[47] with nontrivial topological electronic states[48-50]. This compound hosts a rich array of physical phenomena, such as magnetic Weyl semimetal behavior[51-57], the anomalous Hall effect with large Hall angle[58-60], the anomalous Nernst effect[52,61], flat band orbital magnetism[62-64], giant magneto-orbital responses[65], and unconventional quantum impurity effects[63,64,66]. The cleavage of Co$_3$Sn$_2$S$_2$ can easily produce atomic flat surfaces without apparent surface reconstruction. As such, Co$_3$Sn$_2$S$_2$ provides an excellent platform for investigating correlated topological phenomena at the atomic scale[10,51,67].

Recent advances in scanning tunneling microscopy have substantially enriched our understanding of $Co_3Sn_2S_2$, particularly in identifying surface terminations[68-84], probing magnetism at the atomic scale[71,72,85-91], and elucidating the interplay between electronic correlations and band topology[8,50,59,60,62-64,66,92-94]. Despite these achievements, several key questions remain unresolved: (a) how to reliably identify cleavage surfaces through a decisive method[95]; (b) how to establish direct correspondence between quasiparticle interference patterns and Fermi arcs, and how to establish the nontrivial topology of the edge modes; and (c) how to interpret the behavior of defect-induced quantum states under external fields[66] and spin-polarized tunneling technique[63,64,96-100]. This perspective provides a comprehensive discussion of recent scanning tunneling microscopy-based investigations of $Co_3Sn_2S_2$, underscoring the remarkable progress made in resolving its atomic-scale topological and magnetic properties.

## 2 Identification of the cleavage surfaces

We begin by introducing different methodologies for surface identification and propose imaging designer layer-selective chemical markers as a decisive method. In $Co_3Sn_2S_2$, different cleavage terminations yield markedly distinct electronic structures, making their unambiguous identification crucial. Atomic-step geometry imaging has been introduced as a general methodology for determining surface terminations across six kagome material families, and has been applied[95] for $Co_3Sn_2S_2$. As shown in Fig. 1(a), the crystal structure of $Co_3Sn_2S_2$ (space group $R\bar{3}m$) features robust bonding between the kagome $Co_3Sn$ layers and adjacent S layers. Cleavage preferentially occurs at the weakly bonded Sn interlayer, producing two distinct surfaces: (1) a S-terminated surface decorated with Sn adatoms (red region, Fig. 1(b)), and (2) a Sn-terminated surface exhibiting vacancy defects (blue region, Fig. 1(b)). According to the crystalline geometry, when these two surfaces intersect at a step edge, the Sn-terminated surface forms the upper terrace, while the S-terminated surface forms the lower terrace (Fig. 1(c)). This distinctive geometric feature aids in their identification. In addition, as a natural consequence of unavoidable incomplete cleavage, there will be Sn adatoms on the S surface, as well as Sn vacancies on the Sn surface (Fig. 1(c)). Atomic-resolution imaging (Fig. 1(d)) confirms that Sn adatoms occupy the centers of S hexagonal rings, while Sn vacancies appear at their expected lattice sites.

Despite sharing identical in-plane hexagonal symmetry ($a$ = 5.3 Å), the two terminations exhibit distinct corrugation amplitudes[62]. The Sn-terminated surface displays reduced topographic contrast compared to the S-terminated one, consistent with the larger atomic radius of Sn (1.41 Å) relative to S (1.02 Å). This contrast is further supported by differential conductance spectra and density-of-states calculations (Fig. 1(e)) [63]. Notably, the S-terminated surface hosts a sharp spectral peak at -8 meV below the Fermi level, which is identified as a signature of the kagome-derived flat band in reference to first-principles calculations[62].

To further validate the Sn-termination assignment, ref. 63 employed chemical marker experiments on In-doped $Co_3Sn_2S_2$ (nominally 1% In). Large-scale topography (Fig. 1(f), left) confirms the expected substitution rate, and atomic-resolution images (Fig. 1(f), right) clearly show In atoms substituting into Sn lattice sites. However, pristine $Co_3Sn_2S_2$ contains a non-negligible concentration of native impurities within the Sn layer[101-103], which can obscure dopant interpretations. In addition, the In dopant may also substitute into the $Co_3Sn$ layer, producing distinct defect patterns on the S-terminated surface, a scenario that remains insufficiently investigated. Therefore, we propose imaging layer-selective chemical markers by introducing higher doping levels with near isovalent substitutions, such as Fe for Co, In for Sn, and Se for S, to enable unambiguous layer-specific identification. Such targeted substitution, ideally up to ~10%, can directly link observed topographic features with their crystallographic origin, minimizing ambiguity from native defects. Both the S surface and Sn surface should be imaged with similar tunneling conditions, and eventually six independent topographies on three doped crystals should conclude the surface identification.

On rare occasions, cleavage exposes the $Co_3Sn$ kagome layer (Fig. 1(g))[104,105]. Its surface closely resembles that of the kagome layer in $Fe_3Sn$[96,99,106,107], appearing as a distorted honeycomb lattice due to the short interatomic distances (~2.7 Å) between transition metal sites. In topographic images, the central Sn atoms in this layer manifest as pronounced topographic depressions[107,108]. We should be cautious that, based on the crystal structure, such cleavage is highly unnatural, and it is highly possible that stacking errors in $Co_3Sn_2S_2$ crystals leading to such kagome cleavage surfaces, which deserves future investigation.

The robustness of the imaging atomic step geometry method is further corroborated in the isostructural compound $Ni_3In_2Se_2$ (Fig. 1(h))[95,109]. Cleavage in this material similarly yields Se-terminated surfaces with In adatoms and In-terminated surfaces with In vacancies (Fig. 1(i)). Again, based on the crystalline geometry, when Se surface and In surface meet at a step edge, the upper surface will be the In surface and the lower one will be Se surface. Notably, strong hybridization between Ni 3$d$ and Se 4$p$ orbitals suppresses the Se-derived triangular lattice at positive bias (+200 mV), while revealing kagome-derived electronic states at negative bias (-200 mV). The detection of kagome-like geometry by imaging the Se surface offers an additional validation of the imaging atomic step geometry approach. In other kagome metals featuring two terminations with distinct lattice geometry (such as $KV_3Sb_5$), the atomic step geometry imaging method has also been proven effective[95], which can be independently checked by the different lattice geometries (as hexagonal K lattice and honeycomb Sb lattice). Thus, this method becomes a generic methodology for identifying surfaces of new quantum materials.

Alternative interpretations regarding the surface terminations of $Co_3Sn_2S_2$ have been proposed. For instance, Morali et al.[94] reported that the majority of observed surfaces exhibit a triangular atomic pattern (Fig. 2(a), inset), which they attribute to Sn- and S-terminated surfaces, whereas a smaller subset displays a hexagonal lattice, assigned to Co-terminated regions (Fig. 2(b), inset). In topographic images, the Co atoms form a kagome lattice that appears hexagonal due to each node comprising a triangle of three corner-sharing Co atoms[104,105], consistent with Fig. 1(g). To further distinguish the two triangular-structured surfaces featuring adatoms and vacancies, in contrast to Fig. 1(b)-(d), Morali et al. analyzed the step height (1.72 nm) between terraces. By aligning the upper terrace with the Co layer, they inferred that the lower terrace corresponds to the S-terminated surface characterized by vacancies, whereas the terrace exhibiting adatoms corresponds to the Sn-terminated surface[94]. However, we should be cautious that the measured step height is mixed by the convolution of the local density of states, which is sensitive to the applied bias voltage. In other kagome metals[94], the deviations of step height typically exceed ±0.6 Å, larger than the S-Sn interlayer distance in $Co_3Sn_2S_2$. Thus, step height alone for termination assignment can lead to systematic errors.

Complementing these arguments, Xing et al.[64] used work function measurements and theoretical modeling to assign adatom-decorated surfaces to Sn termination and vacancy-decorated surfaces to S termination (Fig. 2(c)). While supportive, this method is not definitive, as local work function values measured in compounds similarly convolute effect of density of states and are often different from the expected values, and it has not yet become a generical method for surface identification to our understanding. Similarly, Howard et al.[60] used subsurface defect symmetry to support this assignment (Fig. 2(d)-(e)). On vacancy-decorated surfaces, triangular-shaped defects are observed, each spanning three lattice sites and characterized by one vertex appearing notably brighter than the others (Fig. 2(f)). The bright vertex (labeled 1-3 in Fig. 2(e)) is proposed to be Co-site vacancies or substitutional impurities in the underlying $Co_3Sn$ kagome plane due to expected symmetry (Fig. 2(g)). However, a careful comparison of the topographic data presented in refs. 60, 64, and 94 reveals that the adatoms observed in ref. 60 exhibit a "ghosting effect," indicative of tip-induced artifacts, which is absent from the other datasets. This inconsistency underscores the need for caution in interpreting topographic images of local defect geometry, as tip conditions can strongly influence apparent geometrical features of local defects.

At present, the lack of a generally accepted methodology for identifying surface terminations remains a bottleneck in fully understanding the intrinsic properties of $Co_3Sn_2S_2$. It is our perspective that imaging the step edge geometry can be a generic method for surface identification, although it is experimentally challenging and requires searching for such steps, which often takes weeks. Moving forward, we advocate for a chemically controlled approach using imaging designer layer-selective substitutional markers. Unlike methods based solely on topographic contrast or work function measurements, such designer doping targeted at specific atomic planes offers an unambiguous and artifact-resistant method to assign surface terminations with high confidence, thereby enabling more rigorous studies of surface-dependent phenomena in topological kagome systems.

## 3 Flat band orbital magnetism driven by Berry curvature field
In the scanning tunneling microscopy of topological magnet, one approach is to identify unusual spectroscopic features and explore their tunability with perturbations such as magnetic field, then to seek for theoretical understanding. The flat band orbital magnetism in the kagome magnet $Co_3Sn_2S_2$, by unexpected experimental observations, provides an exciting platform for exploring low-energy flat band physics under strong spin-orbit interaction[57,62]. As shown in Fig. 1(e), scanning tunneling spectroscopy reveals a pronounced peak at -8 meV below the Fermi level on the claimed S-terminated surface, while the claimed Sn-terminated surface shows no significant features near the Fermi energy. Comparison with bulk local density of states calculations attributes this sharp peak to a flat band predominantly derived from Co 3*d* orbitals in the kagome layer.

The first principles calculation shows that near the Γ point, there is a spin-polarized kagome flat band in touch with a parabolic band, and spin-orbit coupling will lift their degeneracy[62,110,111] (Fig. 3(a), upper panel). This behavior is captured by a tight-binding model of a single-layer ferromagnetic kagome lattice with nearest-neighbor hopping (Fig. 3(a), lower panel). In the absence of spin-orbit coupling, the lattice hosts a flat band degenerate with a parabolic band at the Γ point, originating from destructive quantum interference among the three sublattices. Ferromagnetism breaks time-reversal symmetry, splitting the spin-degenerate bands. Within each spin sector, spin-orbit coupling lifts the degeneracy and opens an energy gap at Γ. This makes the flat band slightly dispersive and induces a nontrivial Berry phase. The integrated Berry curvature on the flat band yields a Chern number C = 1, elucidating the topological origin of the kagome flat band. The nontrivial Chern topology can manifest in various phenomena, including the anomalous Hall effect[58,112] and the emergence of chiral Weyl fermions[53-56,60,94,113], which are accessible via magnetic field-dependent spectroscopic probes.

The origin of this anomalous flat band magnetism was investigated in ref. 62 through measurements of the field-dependent energy shift of the -8 meV peak. As shown in Fig. 3(b), the peak shifts symmetrically to higher energy under both positive and negative magnetic fields applied along the *c*-axis. Spin-polarized d*I*/d*V* spectra on this surface (Fig. 3(c), claimed otherwise as the Sn surface) exhibit a dominant spin-up component, and linear fitting of the energy shift yields an effective magnetic moment of $m \approx -2.9\mu_B$, consistent with earlier studies (Fig. 3(d)) [64].

This behavior is quite different from the conventional Zeeman expectations[62,114]. For a spin-polarized state, it can be expected that opposite field directions will shift the state to opposite energy directions. So the first observation that the state always shifts to the same energy direction, tells us that the spin polarization follows the magnetic field direction, as indeed its coercivity field is only 0.3 T, much smaller than the step size of the applied field, as 2 T. We call this phenomenon the magnetization polarized Zeeman effect, uniquely occurring in magnetic systems with spin-polarization. Secondly, the spin-up state is expected to shift to lower energies with increasing magnetic field, again in contrast with the observations. This indicates an effective negative magnetic moment. The extracted moment (~ -3$\mu_B$) by analyzing the shift rate as a function of fields far exceeds the expected spin contribution (~ +1$\mu_B$), suggesting a dominant orbital origin. This points to a Berry curvature--induced orbital magnetization intrinsic to the topological flat band[115-117].

Theoretical analysis (Fig. 3(a),(e)-(f)) confirms that the negative moment arises from spin-orbit-induced orbital magnetism tied to the nontrivial Berry curvature of the flat band[118,119]. First-principles calculations and tight-binding models reveal a substantial negative orbital moment localized within the flat band, even though the system is globally dominated by the spin magnetism. In the spin-orbit coupled magnetic kagome model, the orbital magnetic moment of a Bloch band is given by the modern theory of Berry curvature-induced magnetism[118,119]: $m_{\text{orbital}}(k) = Im < \partial_k u_a | \times [H^{eff}(k) - \epsilon_a(k)] | \partial_k u_a >$ (*k* is momentum, $u_a$ is the eigenstate, $H^{\text{eff}}$ is the Hamiltonian, and $\epsilon_a(k)$ is the band dispersion). Under the *k·p* approximation, this yields an analytical expression $m_{\text{orbital}}(k) = -\frac{k^2 t^2 \lambda}{k^4 t^2 + 48\lambda^2}$ for the flat band (*t* is the nearest neighbor hopping integral and *λ* is the spin-orbit coupling constant). This localized orbital diamagnetism underscores the role of spin-orbit coupling in shaping the topological and magnetic responses of kagome bands. These findings provide key insights into emergent orbital magnetism in correlated kagome systems[34,120-124] and offer a new avenue to flat band related orbital magnetism other than moiré systems such as magic-angle twisted graphene[119,125,126]. In the magic-angle twisted graphene, when the flat band is gated through the Fermi level, it exhibits an emergent Hubbard splitting behavior[127-131]. The magnetic field driven kagome flat band shift is also toward the Fermi level, thus it is highly interesting to further increase the magnetic field to drive the kagome flat band to be at the Fermi level to search for related correlated topological phenomena.

**4 Detecting the Fermi arc and topological edge state**
Besides the experimental observation driven approach, another valuable approach for scanning tunneling microscopy of topological magnet is to visualize outstanding predictions based on first-principles calculation on the consequence of quantum topology. In this regard, Co₃Sn₂S₂ has been predicted to feature magnetic Weyl topology[55,58,59,94], which can produce Fermi arcs[54] and topological edge modes[53] that can be detected by surface-sensitive probes. As shown in Fig. 4(a), initial theoretical calculations reveal that on the Sn-terminated surface, the Fermi arc-associated band extends from the Weyl point energy ($E_W$) up to the Fermi level ($E_F$), offering an energy window for probing arc-like features[54]. In contrast, on the S-terminated surface, the Fermi arc-related band lies energetically above the Weyl point $E_W$, suggesting that the contribution of these states to spectroscopic signals, including quasiparticle interference patterns, can differ depending on surface termination.

The theoretically computed quasiparticle interference maps in ref. 54, which account for all possible surface-state scattering channels, highlight distinct interference patterns between Sn- and S-terminated surfaces. However, due to the complexity arising from overlapping scattering processes, directly correlating initial theoretical predictions with experimental observations is challenging. In experiment, the claimed Sn-terminated surface exhibits sharp, polygonal quasiparticle interference features at low energies (~ -10 meV, Fig. 4(b), top), including a honeycomb-like pattern centered around Bragg peaks and linear features along the Γ-M direction. These linear segments connect adjacent corners of the hexagon and are proposed to reflect quasi-nesting conditions between adjacent triangular pockets at K and K'. The Fermi-arc segments involved in these scattering process bridge pairs of Weyl nodes within the Brillouin zone[94]. On the claimed S-terminated surface, experimental Fourier-transformed d$I$/d$V$ maps at 100 meV (Fig. 4(b), bottom) reveal broader quasiparticle interference features along the Γ-M direction, accompanied by a central, flower-like pattern. These features are attributed to interband scattering involving dominant surface states that may hybridize with Fermi-arc states near the projected Weyl nodes. Importantly, the connectivity of Fermi arcs in the first-principles calculation depends on the surface, as illustrated in Fig. 4(c): samples with Sn and Co terminations on opposite surfaces exhibit interlayer Fermi-arc transport that requires electrons to traverse six arcs and the bulk six times in a full cycle. In contrast, previous studies considered surfaces with identical connectivity, where electrons oscillate only through a single Fermi arc on each of the two surfaces[132,133].

From the broader perspective of studying quantum materials with scanning tunnelling microscopy, the termination-dependent quasiparticle interference patterns are generically consistent with theoretical expectations, even without invoking topological effects. And if different surfaces exhibit similar quasiparticle interference patterns, such a robust signal can often be suggestive of certain topological protection. Additionally, differences exist among various theoretical studies. For example, refs. 94 and 54 report distinct quasiparticle interference predictions for the same surface termination, likely due to differences in the surface-state scattering channels included in their respective models. A deeper understanding of these discrepancies can be achieved by examining the surface-state band structure. As shown in the top image of Fig. 4(d), calculations from ref. 55 place the Fermi arcs near the K' point, consistent with those in ref. 94. Each arc connects a pair of Weyl nodes of opposite chirality and merges into the bulk Fermi pockets near the M' point. An alternative and more recent perspective (Fig. 4(d), bottom), based on similar *ab initio* calculations from overlapping authors, suggests that the apparent arc near K' may largely correspond to a trivial surface state encircling the K point (yellow arrow), while a nascent topological arc (white arrow, with much weaker intensity) connects projected Weyl cones. The coexistence of trivial and topological states complicates the quasiparticle interference response, as multiple scattering channels, including scattering between trivial states, scattering between trivial and topological arcs, and scattering between arcs, simultaneously contribute, thereby requiring careful analysis to disentangle their individual signatures. Therefore, it is our perspective that in such systems, regardless of the surface identification issue, it is far from clear that we can directly extract the Fermi arc signal from the quasi-particle interference data. An ideal case for establishing the Fermi arc should be that we put enough constraints on the origin of the quasi-particle interference signals by experiment alone. This is unfortunately too challenging to realize in this correlated system, and we look for future advances in addressing such challenge. We also comment that it is a generic case that the trivial surface states coexist with topological Fermi arcs in many Weyl systems by first-principles calculations. Then, even for the direct momentum space measurement by angle-resolved photoemission spectroscopy, there is room that researchers can assign the observed surface states with either topological Fermi arcs or trivial surface states by fine-tuning the surface relaxation parameters in the first-principles calculations when the position of the Weyl cones is not accurately determined by experiment. A strong constraint would be to experimentally build up the momentum connectivity between Weyl cones and Fermi arc tips.

To complement this momentum-space analysis, real-space studies near step edges offer additional insight into the boundary modes. $Co_3Sn_2S_2$ with atomic thickness is theoretically predicted to host a chiral edge state[53]. This is because at the atomic thickness limit, $Co_3Sn_2S_2$ becomes a Chern insulator, and with increasing thickness, the Chern energy gap may close at certain $k_z$ points, leading to magnetic Weyl fermions. It can be expected that the surface step edge of a bulk crystal may also host edge states, despite its coupling with the bulk substrate. Furthermore, the properties of the edge states are dependent on the nature of the terminations. Although both Sn- and S-terminated surface exhibit nontrivial edge state, the Sn termination only has 3 chiral edge states with Chern number of 3 for the trivial dangling bond states (Fig. 4(e), left), while Chern number is 6 for S termination with semimetallic state (Fig. 4(e), right)[53]. Figure 4(f) shows the residual local density of states along the $y$-direction after removing lattice-induced oscillations and background, revealing standing-wave patterns near the step. The open circles mark wave peaks, and the cross-sectional profile at x = 0 nm (light blue) shows clear modulation of the local density of states at the edge. In Fig. 4(g), a linear relationship is proposed between energy and inverse wavelength for the quantum well-like states, yielding a dispersion velocity of approximately $5\times10^4$ m/s. The inset displays $dI/dV$ maps showing quantum bound states with varying nodal structures across energies, confirming their positive dispersion. The linear relationship is most crucial for connecting to the underlying topology. Based on the data error bar, the linear relationship only strongly distinguishes from the parabolic relationship (topologically trivial) on the small wave vectors. It would be highly valuable to either reduce the systematic data error bar or extend the data to smaller vectors in the future to fully establish the topological nontrivial feature of the edge modes and their connection to Weyl topology. In addition, the topological edge mode can often correspond to a topological energy gap in the bulk, it would be desirable to build up such bulk-boundary correspondence by comprehensive spectroscopic measurements[120,134].

**5 Defect bound states and their connection to flat band magnetism and Weyl topology**
Localized impurity states provide a powerful alternative avenue for probing spin-orbit-driven phenomena in $Co_3Sn_2S_2$. The challenging spin-polarized scanning tunneling microscopy technique has been applied to resolve the local magnetism[8,79,96,135-137]. A systematic spin-polarized scanning tunneling microscopy study was performed using a Ni tip under controlled magnetic fields ($\pm0.1$ T to $\pm0.5$ T) to probe the magnetization of the claimed individual In-induced impurity states[63]. For the spin-polarized tunneling experiment on a ferromagnetic system, it is our understanding that the most crucial experimental setup is to control the spin orientations of the magnetic tip and ferromagnetic sample separately. In this setup, $\pm0.1$ T is used to flip the spin of the soft ferromagnet Ni tip featuring a much smaller coercivity field, and $\pm0.5$ T is used to flip the magnetization of $Co_3Sn_2S_2$ featuring a coercivity field of 0.3 T. By sequentially reversing the tip and sample magnetizations, the impurity resonance was unambiguously identified as spin-down polarized, indicating its anti-aligned with the bulk ferromagnetic order (Fig. 5(a)). As the In impurity is nonmagnetic, the appearance of spin-polarized resonances demonstrates that nonmagnetic scattering in a magnetic host can generate localized magnetic responses. This correlation effect can be well captured by T-matrix calculations that incorporate local potential scattering in a spin-orbit coupled kagome lattice[63].

Subsequent high-field measurements ($|B| = 8$ T) revealed a Zeeman energy shift ($\Delta E \approx +2.2$meV) proportional to the field magnitude (Fig. 5(b)), corresponding to an effective magnetic moment of approximately $-5\mu_B$. This value, far exceeding the expected spin-only contribution, reflects a substantial orbital component arising from spin-orbit hybridization between the nonmagnetic In impurities and the Co $3d$ orbitals in the kagome lattice. First-principles calculations support these findings, locating the resonance within the spin-down gap (inset, Fig. 5(b)), and confirming its origin as a localized perturbation in the spin-polarized electronic structure. This pronounced diamagnetic response underscores the critical role of spin-orbit coupling in enhancing impurity-induced orbital moments. Experimentally, we find that the orbital magnetic moment of the flat band is approximately $-4\mu_B$, based on a total measured moment of $-3\mu_B$ and an estimated spin contribution of $+1\mu_B$. Similarly, the In impurity resonance shows a $-5\mu_B$ moment, which, after accounting for a $-1\mu_B$ spin contribution, yields an orbital moment also around $-4\mu_B$. This numerical similarity suggests a potential underlying link between the impurity resonance (local flat band) and the kagome lattice flat band. While the orbital magnetism of the lattice flat band is well understood within a topological framework (Berry curvature), a rigorous theory for the impurity case remains an open question.

Furthermore, coupled impurity states reflect the spin-orbit driven topology. For example, three impurities arranged in a triangular ($C_{3v}$) geometry exhibit hybridized bound states in the tight-binding model, with one single state at energy $2t$ and two degenerate eigenstates both at energy $-t$. The spin-orbit coupling breaks the mirror symmetry and lifts the two degenerate levels and modifies their energy into $-t\pm\sqrt{3}\lambda$. Such physics resembles the splitting of magnetic nodal lines in producing Weyl fermions. Scanning tunneling microscopy measurements revealed hybridization between In impurities (Fig. 5(c))[63]. For isolated impurities, a single sharp resonance (~ -270 mV) is localized at the defect site (red line, Fig. 5(c)). As the distance between two impurities decreases, this state splits into bonding ($\sigma$, -315 mV) and antibonding ($\sigma^*$, -240 mV) orbitals (blue line), confirmed via spatially resolved d$I$/d$V$ maps. Strikingly, a trimer of impurities arranged with $C_{3v}$ symmetry resolves into three distinct levels: a bonding orbital (-310 mV) and two non-degenerate antibonding states (-220 mV and -170 mV) (violet line). This splitting deviates from the expected two-fold degeneracy due to mirror symmetry protection. While the geometry predicts degenerate $\sigma^*$ states, the atomic spin-orbit interaction lifts this degeneracy (Fig. 5(d), upper panel). The resulting ~50 meV spin-orbit gap is corroborated by angle-resolved photoemission spectroscopy measurements (Fig. 5(d), lower panel)[9,57]. This impurity-state hybridization closely resembles the bulk splitting of topological nodal lines (protected by crystalline mirror symmetry) in $Co_3Sn_2S_2$, establishing a link between atomic-scale spin-orbit impurity phenomena and the bulk spin-orbit coupling related topology.

Independent spin-polarized scanning tunneling microscopy studies have revealed spin-polarized bound states at claimed native S vacancies[64]. The d$I$/d$V$ spectra (Fig. 5(e)) show pronounced magnetic contrast in the bound-state energy range (-350 to -280mV), with a dominant spin-down character. Figure 5(f) demonstrates the effect of tip spin orientation under ±0.6 T fields: the resonance peak at -283mV is suppressed under a spin-up tip and restored when the spin-down configuration is reinstated, which is proposed to demonstrate the presence of localized magnetic polarons associated with S vacancies and the robustness of their spin-orbit nature.

These measurements also constrain the coercivity of the system, which is different from the bulk and previous studies. Prior studies report a bulk coercive field of 0.3 T[64]. If we take this value, when fields exceeding ±0.6 T are applied, both tip and sample magnetizations are expected to be flipped or polarized. However, the sample or the defect's spin direction is assumed to be unaffected, suggesting that the sample's coercive field can be larger than 0.6 T. In addition, the magnetization polarized Zeeman effect of this impurity state (Fig. 5(g)) demonstrates that 1T field readily flips the spin of the sample, otherwise, the state will shift to the opposite direction when reversing the fields. To our understanding, if there is no systematic error, these sets of data can only be reconciled if the samples coercivity field is just between 0.6 T and 1T. In future spin-polarized experiments, it would be highly useful to progressively determine both coercivities of the tip and sample by reducing the magnetic field steps, which have been systematically lacking in both the claimed In defect case and claimed S vacancy case.

The bound-state energy shift ($\Delta E$) under out-of-plane field ($B_z$) displays an anomalous linear Zeeman response. Unlike conventional spin magnetic moments, which shift asymmetrically with field polarity, the observed $\Delta E$ increases symmetrically with field magnitude (Fig. 5(g)), signaling a dominant orbital magnetization component. This behavior can arise from orbital diamagnetic currents associated with nontrivial Berry curvature, a signature of topological materials. The resulting topological magnetoelectric response is intimately tied to phenomena such as the anomalous Hall effect and orbital magnetism[64]. The evolution of the impurity magnetic moment ($m$) with vacancy cluster size (N) (Fig. 5(h)) reveals a smooth crossover from localized to itinerant magnetism. For small vacancies (e.g., N = 3), m is positive (~ +1.55 $\mu_B$), indicating spin-polarized bound states akin to proposed magnetic polarons. As N increases, m gradually becomes negative, reaching ~ -3.28 $\mu_B$ for large clusters (N → ∞). This transition reflects a shift toward orbital-dominated, delocalized magnetism originating from the extended claimed S hexagonal lattice, which is proposed to provide evidence for the emergence of itinerant spin-orbit polarons[66].

Together, these scanning tunneling microscopy studies establish that atomic defects in $Co_3Sn_2S_2$ host strongly spin-orbit-coupled impurity states that manifest as orbital-dominant magnetic resonances. These spin-orbit resonances are tunable through controlled defect clustering, enabling manipulation of their energy levels and magnetic character. The coupling between these localized states, topological surface modes, and underlying magnetoelastic interactions provides a promising pathway for defect-engineered control of magnetism and topology in kagome quantum materials.

# 6 Conclusion

Scanning tunneling microscopy studies of $Co_3Sn_2S_2$ have yielded critical insights into surface terminations, local topological states, and electron correlation effects at the atomic scale. The distinct Sn- and S-terminated surfaces exhibit characteristic defect configurations and electronic structures. We propose to use the generic step edge geometry imaging for surface identification, and we advocate for the adoption of comprehensive, designer layer-selective chemical markers as a robust and unambiguous strategy for surface identification in $Co_3Sn_2S_2$ and related kagome systems. Crucially, the observation of kagome-derived negative flat band magnetism underscores the dominant role of Berry curvature induced orbital moments, which manifest in ways that markedly deviate from conventional Zeeman behavior. We further propose to use a larger magnetic field to drive the kagome flat band to Fermi level to uncover correlated topological phenomena. The magnetic Weyl topology predicts diverse Fermi arc behaviors and topological edge modes. Different quasi-particle interference patterns observed on different terminations have been proposed to support the diverse Fermi arc connectivity. Edge modes observed with near-linear dispersion are also proposed to support the nontrivial band topology. Further improvements to enhance the experimental constraints on topology are discussed. Additionally, spin-polarized scanning tunneling microscopy has enabled the direct visualization of quantum impurity states and proposed magnetic polarons, revealing the tunability of orbital magnetization through defect engineering at the atomic scale. The coupling between multiple impurity states is deeply connected to flat band magnetism and the Weyl topology. We further propose to sweep the magnetic field more systematically to demonstrate the separate spin flip process of the sample and tip in the spin-polarized tunneling experiment.

We also note that recent discovery[138,139] of chiral phonons in $Co_3Sn_2S_2$ shows that magnetic, electronic and lattice degrees of freedom are strongly coupled. In the kagome lattice vibration model[105], the deformation of a hexagonal ring (inner six atoms in a kagome pattern) rotating clockwise or anti-clockwise, without displacements of the outer atoms, are localized eigenmodes for each ring. Jointly, these localized vibrations form the phonon flat-band. Due to overall inversion symmetry of the crystal, these "pre-formed" chiral phonons are degenerate with the antichiral partners. Only by coupling to magnetic degrees of freedom, this degeneracy gets lifted. In turn, the chiral flat phonon band can serve as an Einstein mode that couples strongly with the electronic band, leading to substantial band renormalizations. This Fermion-Boson many-body interplay deserves future investigation via quasiparticle interference techniques.

Together, these discoveries and discussions deepen our fundamental understanding of the interplay between topology, magnetism, and electron correlation in kagome materials. More broadly, they open promising avenues for precise control of topological states and spin-orbit coupling phenomena, offering an exciting foundation for the design of next-generation spintronic and quantum information devices.

**Acknowledgments**
We acknowledge the helpful discussion with Songtian S.Zhang, Lei Shan, Berthold Jaeck and Hermann Suderow.

**Authors' contributions**
G.L. and W.S. wrote the draft of this manuscript in consultation with J.X.Y.; T.N., M.Z.H. and H-B.D. revised and provided suggestion to the manuscript. All the authors read, revised, and approved the final version.

**Funding**
J.X.Y. acknowledges the support from the National Key R&D Program of China (Nos. 2023YFA1407300, 2023YFF0718403), the National Science Foundation of China (Nos. 12374060), Guangdong Provincial Quantum Science Strategic Initiative (GDZX2401001).

**Availability of data and material**
The data that support the findings of this study are available from the corresponding author upon reasonable request.

**Competing interests**
The authors declare no competing interests.

## Figures and captions

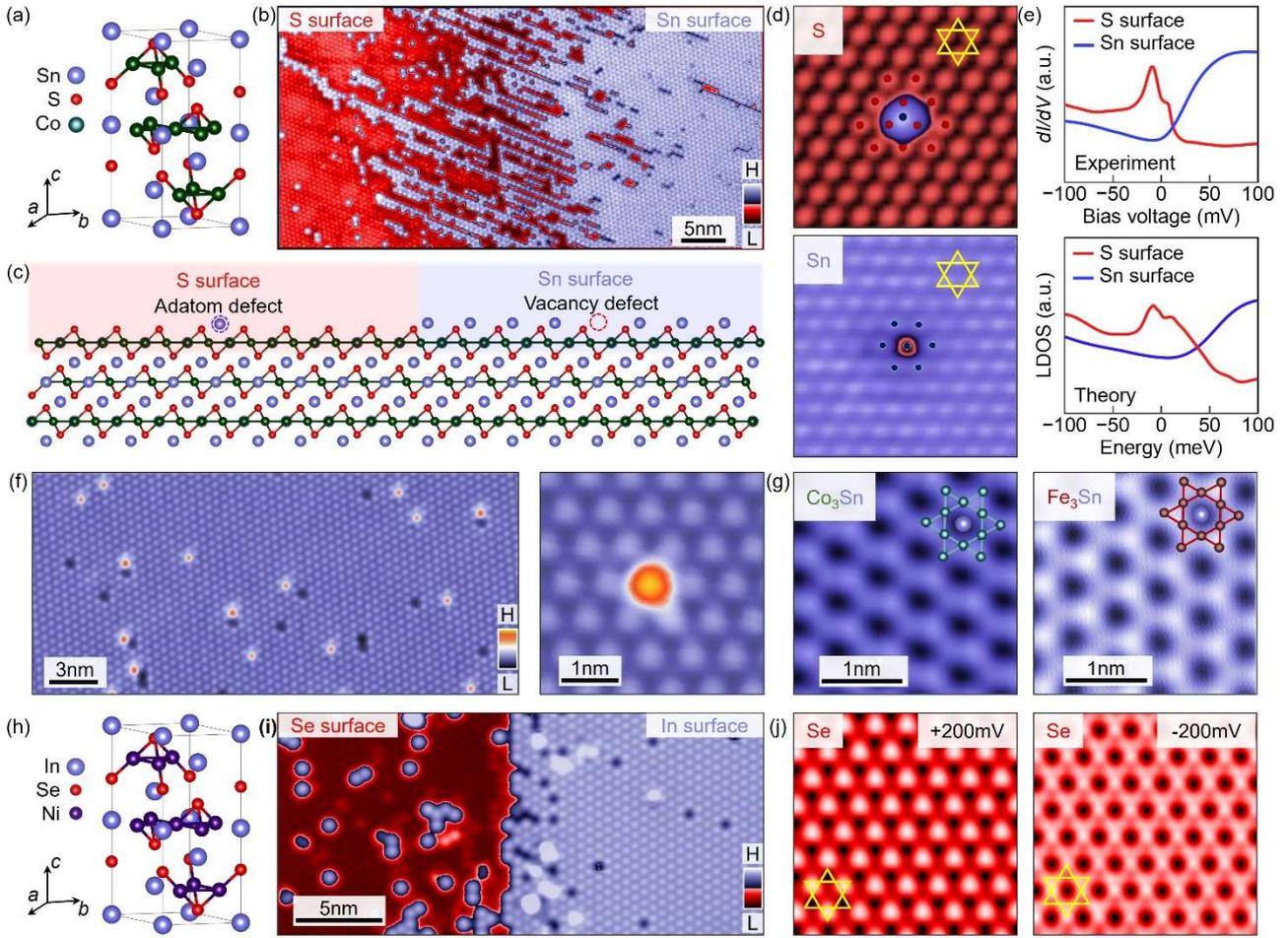

**Figure 1** *Surface termination identification by imaging atomic step geometry in $Co_3Sn_2S_2$ and $Ni_3In_2Se_2$.* (a) Crystal structure of $Co_3Sn_2S_2$. (b) Topographic image of the boundary between S-terminated (red) and Sn-terminated (blue) surfaces. (c) Schematic of cleavage-induced defects: Sn adatoms on the S surface (left) and Sn vacancies on the Sn surface (right). (d) Atomic-resolution scanning tunneling microscopy images of adatom (top) and vacancy (bottom) defects. (e) Experimental d$I$/d$V$ spectra (top) and theoretical density of states (bottom) for S- and Sn-terminated surfaces. (f) Sn-terminated surface of 1% In-doped $Co_3Sn_2S_2$: Large-scale topography (left) and atomic-resolution image of an In substitutional defect (right). (g) Comparative scanning tunneling microscopy images of $Co_3Sn$ kagome surfaces in $Co_3Sn_2S_2$ (left) and $Fe_3Sn$ surfaces in $Fe_3Sn_2$ (right), both exhibiting distorted honeycomb morphologies. (h) Crystal structure of $Ni_3In_2Se_2$, isostructural with $Co_3Sn_2S_2$. (i) Boundary between Se-terminated (red, adatom defects) and In-terminated (blue, vacancy defects) surfaces in $Ni_3In_2Se_2$. (j) Bias-dependent surface states of $Ni_3In_2Se_2$: Triangular lattice at $V = +200$ mV (left) and kagome-derived electronic states at $V = -200$ mV (right), induced by Ni 3$d$-Se 4$p$ orbital hybridization. Figures adapted from refs. 59,62,63,95.

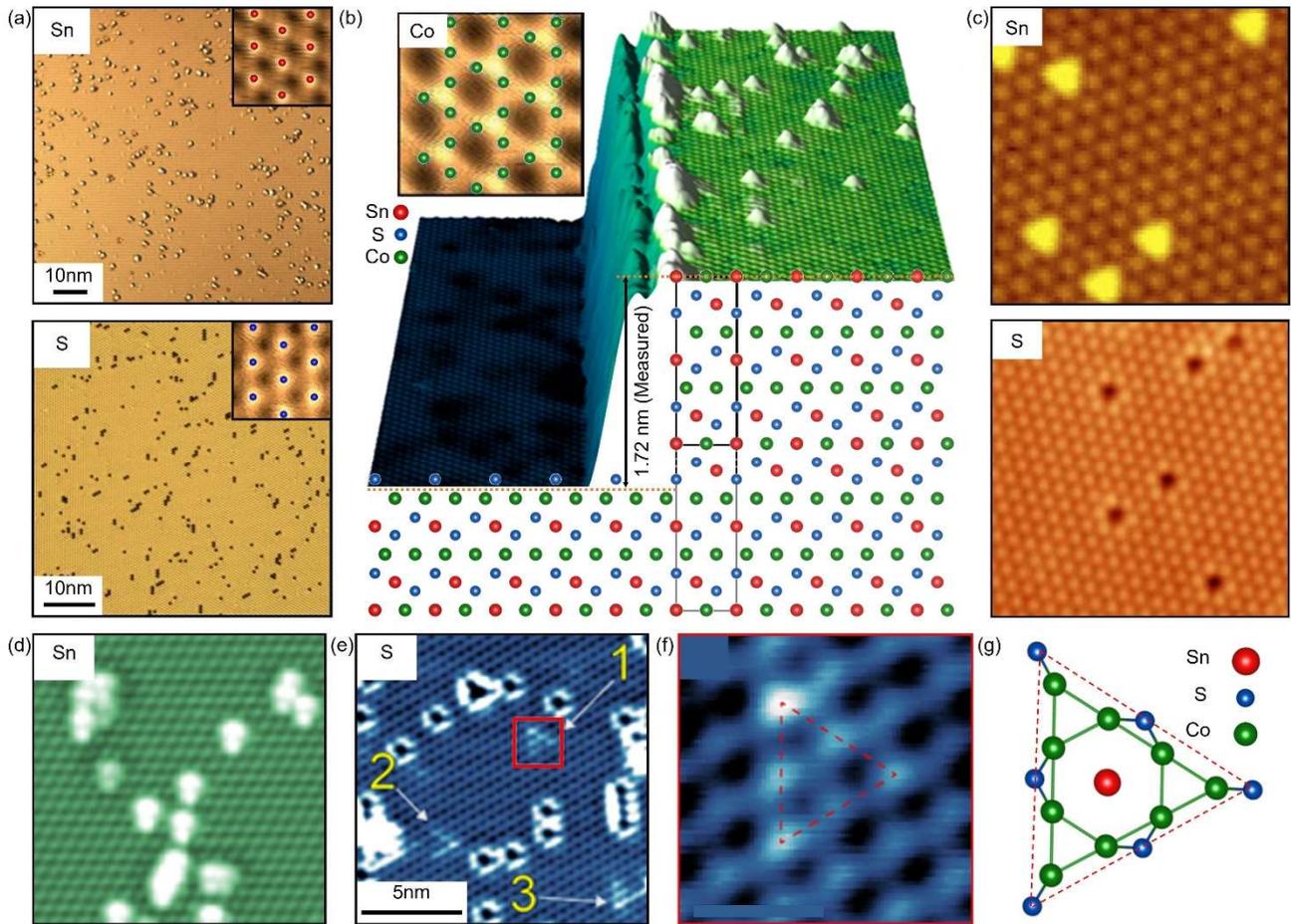

**Figure 2** *Surface termination identification by step height and unknown impurities.* (a) Top topographic image of the Sn-terminated (001) surface displays adatoms and subsurface impurities, and the bottom S-terminated (001) surface reveals S vacancies, respectively. (b) 3D topographic image of the $Co_3Sn_2S_2$ surface, highlighting a step between the Co- and S-terminated regions. The superimposed crystal structure of $Co_3Sn_2S_2$ illustrates the cleavage planes, where Co, S, and Sn atoms are represented in green, blue, and red, respectively, with the black solid line marking the unit cell. Inset: Atomically resolved Co surfaces. (c) Topographic image of the Sn and S-terminated (001) surface. (d) Topography image highlighting adatoms as characteristic features of the Sn surface. (e) Topography image highlighting vacancies as characteristic features of the S surface, with three distinct defects labeled as 1, 2, and 3, corresponding to bright vertices pointing upward, rightward, and leftward, respectively. (f) The region within the red box in (e) provides a zoomed-in view of the triangular defect; a red dashed triangle is included for comparison with (g). (g) Schematic representation of the triangle in (f), depicting the top-layer S atoms and the underlying $Co_3Sn$ layer. Figures adapted from refs. 60,64,94.

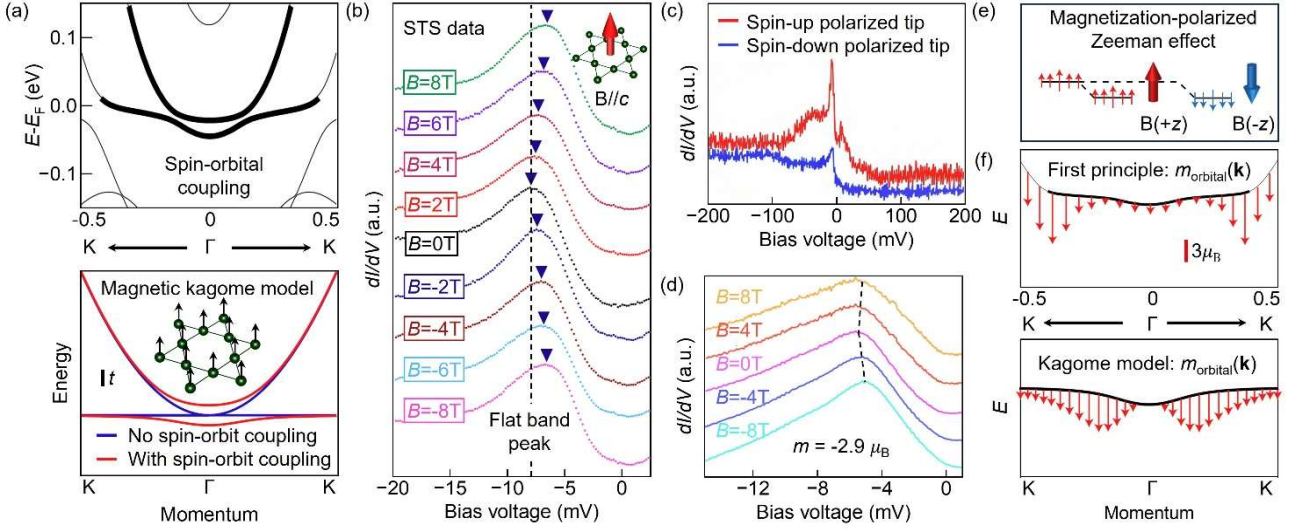

**Figure 3** *Flat band orbital magnetism driven by Berry curvature field.* (a) Upper panel: First-principles spin-resolved band structure of $Co_3Sn_2S_2$ at $k_z = 0$, calculated with spin-orbit coupling. Lower panel: Tight-binding model of a single-layer ferromagnetic kagome lattice (inset), showing calculated band dispersions with (red) and without (blue) spin-orbit coupling. (b) Magnetic field ($B/\!/c$) dependence of the flat band peak on the S-terminated surface. The peak shifts symmetrically to higher energies under both $+B$ and $-B$, with identical magnitude. (c) Spin-polarized d$I$/d$V$ spectra on S surface, showing a larger response when measured with a spin-up polarized tip. (d) Comparison with prior studies: Field-dependent energy shift (slope 165 μeV/T = $2.9\mu_B$) on the identified S surface, consistent with our results. (e) Schematic of the magnetization-polarized Zeeman effect. Spin alignment under $\pm B$ induces identical energy shifts, independent of field direction. (f) Upper panel: First-principles orbital magnetic moment ($m_{orbital}$) distribution (red arrows) along the flat band. The red bar denotes $m_{orbital}$ magnitude. Lower panel: Orbital magnetism (arb. units) derived from the kagome lattice model, with $m_{orbital}$ directionality (red arrows). Figures adapted from refs. 62,64.

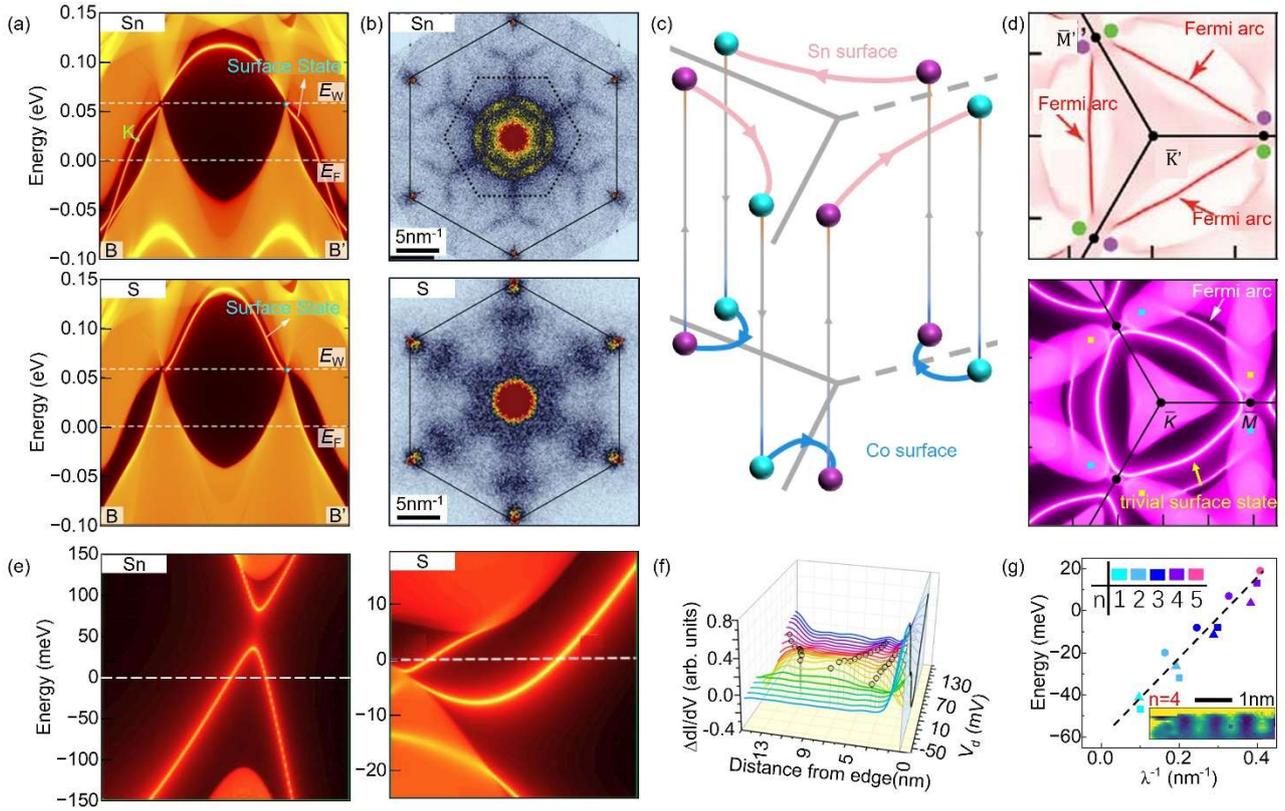

**Figure 4** *Detecting Fermi arc and topological edge state.* (a) Energy dispersion for both Sn- and S-terminated surfaces along a *k*-path that intersects a pair of Weyl points connected by a Fermi arc (B and B' point can be found in ref. 54). (b) Fourier transform of two d$I$/d$V$ maps obtained from different surfaces, revealing sharp quasiparticle interference patterns. (c) Schematic representation of Sn- and Co-terminated surfaces, illustrating their respective intra- and inter-Brillouin-zone Weyl node connectivity, along with the semiclassical electron magnetotransport trajectory (arrows). (d) Calculated Fermi surface including contributions from both bulk and surface states. Magenta and green dots denote Weyl points of opposite chirality, while are indicated by red arrows (top). *Ab initio* calculation of the surface density of states at the Fermi energy for the Sn-terminated surface, incorporating spin-orbit coupling. Weyl points with positive and negative chirality are marked by yellow and cyan squares, respectively (bottom). (e) Energy dispersion of the edge states for Sn surface (left) and S surface (right). (f) Residual local density of states, with lattice-induced waves and an averaged background filtered out, shown at selected bias voltages. (g) Energy dispersion as a function of inverse wavelength ($\lambda^{-1}$) for states identified on three $Co_3Sn$ terraces. Squares represent the states shown in the figure, while circles and triangles correspond to states from two other terraces. The dashed line serves as a visual guide. Inset: density of states maps at specific energies corresponding to the n = 4 quantum well like state, where color represents DOS intensity. Figures adapted from refs. 50,54,55,60,94,113.

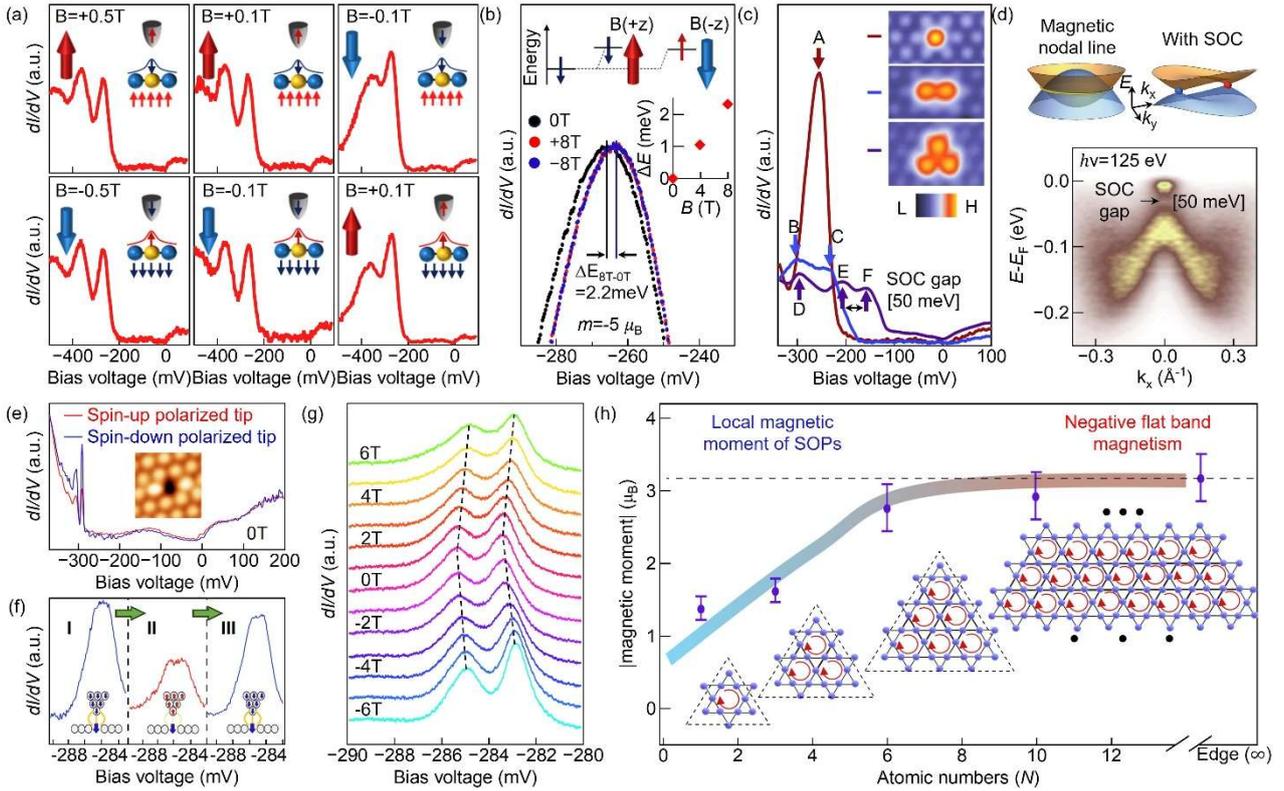

**Figure 5** *Defect bound states and their connection to flat band magnetism and Weyl topology.* (a) Spin-polarized tunneling spectroscopy of the impurity state under weak magnetic fields ($B_z = \pm 0.1 \sim 0.5$ T), revealing spin-down polarization (anti-aligned with bulk magnetization). Insets: Schematics of tip and impurity spin configurations. (b) Magnetic field dependence ($B_z = \pm 8$ T) of the impurity state, showing a magnetization-polarized Zeeman shift (slope: 0.275 meV/T, $m = -5\mu_B$). Inset: Linear energy shift versus $|B_z|$, independent of field direction. (c) Splitting of impurity-bound states for interacting impurities. Two neighboring impurities exhibit spin-orbit-coupled antibonding states ($\sigma^*$) with a 50 meV splitting. (d) Upper panel: Schematic of a spin-orbit-split magnetic nodal line (yellow ring) generating Weyl points (red/blue spheres). Lower panel: angle-resolved photoemission spectroscopy measured band structure showing a 50 meV spin-orbit coupling-induced gap, consistent with STS splitting in (c). (e) Spin-polarized d$I$/d$V$ spectra at a single Sn vacancy, showing spin-down majority (tip magnetization: ↑ / ↓ ). Inset: Atomic-resolution scanning tunneling microscopy image of the vacancy $V$ = -400 mV, $I$ = 200 pA). (f) Reproducible spin-flip spectroscopy at the Sn vacancy by applying $\pm 0.6$ T, confirming spin-polarized tunneling. (g) Magnetic field dependence of the vacancy-bound state, revealing a negative effective moment ($m = -3.28\mu_B$) from anomalous Zeeman shifts. (h) Evolution of magnetic moments with vacancy size (N: atomic number). Localized moments transition to itinerant negative magnetism (red arrows: Berry curvature-induced orbital currents). Colored curve: Gradient line represents curve of best fit. Figures adapted from refs. 9,57,63,64,66.